\documentclass[twocolumn,aps,pra,notitlepage,superscriptaddress,longbibliography]{revtex4-1}
\usepackage[colorlinks, linkcolor = ForestGreen, anchorcolor = blue, citecolor = magenta]{hyperref}
\usepackage{amsmath,amssymb,amsthm,mathtools,mathrsfs}
\usepackage[table,dvipsnames]{xcolor}
\usepackage{tikz}
\usepackage{adjustbox}
\usepackage{dsfont}
\usepackage[normalem]{ulem}
\usepackage{enumerate}
\usepackage[all]{xy} 

\theoremstyle{plain}
\newtheorem{theorem}{Theorem}
\newtheorem{lemma}{Lemma}
\newtheorem{proposition}{Proposition}

\theoremstyle{definition}
\newtheorem{definition}{Definition}

\newtheorem{example}{Example}

\newcommand{\bd}{\begin{definition}}
\newcommand{\ed}{\end{definition}}
\newcommand{\bt}{\begin{theorem}}
\newcommand{\et}{\end{theorem}}
\newcommand{\bn}{\begin{proposition}}
\newcommand{\en}{\end{proposition}}
\newcommand{\be}{\begin{equation}}
\newcommand{\ee}{\end{equation}}
\newcommand{\blem}{\begin{lemma}}
\newcommand{\elem}{\end{lemma}}
\newcommand{\bx}{\begin{example}}
\newcommand{\ex}{\end{example}}
\newcommand{\bprf}{\begin{proof}}
\newcommand{\eprf}{\end{proof}}

\newcommand{\ket}[1]{|{#1}\rangle}
\newcommand{\bra}[1]{\langle{#1}|}
\newcommand{\braket}[2]{\langle{#1}|{#2}\rangle}

\DeclareMathAlphabet{\mathpzc}{OT1}{pzc}{m}{it} 
 \DeclareFontFamily{OT1}{pzc}{}
 \DeclareFontShape{OT1}{pzc}{m}{it}{ <-> s*[1.2] pzcmi7t }{}
 \DeclareMathAlphabet{\mathpzc}{OT1}{pzc}{m}{it}
 
\newcommand{\Ad}{\mathrm{Ad}}
\def\R{{{\mathbb R}}}
\def\C{{{\mathbb C}}}

\newcommand{\Tr}{\operatorname{Tr}}

\def\O{\mathscr{O}}

\def\C{\mathbb{C}}

\def\R{\mathbb{R}}

\begin{document}									
\preprint{APS/123-QED}

\title{On Lorentzian symmetries of quantum information}

\author{James Fullwood}
\affiliation{School of Mathematics and Statistics, Hainan University, Haikou, Hainan Province, 570228, China}
\author{Vlatko Vedral}
\affiliation{Clarendon Laboratory, University of Oxford, Parks Road, Oxford OX1 3PU, United Kingdom}
\author{Edgar Guzm\'{a}n-Gonz\'{a}lez}
\email{edgar.guzman@hainanu.edu.cn}
\affiliation{School of Physics and Optoelectronic Engineering, Hainan University, Haikou, Hainan Province, 570228, China}

\date{\today}

\begin{abstract}
A foundational result in relativistic quantum information theory due to Peres, Scudo, and Terno, is that von~Neumann entropy is not Lorentz invariant. Motivated by the ``It from Qubit" paradigm, here we show that Lorentzian symmetries of quantum information emerge naturally in a pre-spacetime setting, without any reference to external variables such as position or momentum. In particular, we derive the natural action of the restricted Lorentz group $\text{SO}^+(1,3)$ on the internal degrees of freedom of a single qubit from a simple, information-theoretic principle we refer to as \emph{preservation of linear entropy}. It is then shown that the Lorentz invariance of the linear entropy of a relativistic qubit is a special case of a much more general phenomenon, namely, that any spectral invariant of an operator we term the `$W$-matrix' is an $\text{SL}(2,\C)^{\otimes n}$ invariant scalar. Consequently, the linear $n$-partite quantum mutual information is shown to be an $\text{SL}(2,\C)^{\otimes n}$ invariant for all $n$-qubit states. Finally, we show that the correlation function associated with a pair of qubits in the singlet state yields the Minkowski metric on the space of qubit observables, whose symmetry group is the full Lorentz group $\text{SO}(1,3)$. In accordance with the ``It from Qubit" paradigm, our results thus establish the natural emergence of relativistic spacetime structure from intrinsic properties of quantum information.
\end{abstract}

	\maketitle

Traditionally, investigations of relativistic aspects of quantum information proceed by embedding a system of qubits into a fixed spacetime background. In such a context, Peres, Scudo, and Terno showed that when one considers the Lorentz group acting on the momentum degrees of freedom of a qubit, the von~Neumann entropy has no invariant meaning~\cite{Peres_2002}. Such a result essentially follows from the absence of finite-dimensional unitary representations of the Lorentz group, which necessitates a coupling between spin and momentum that renders the von Neumann entropy observer-dependent.

However, the assumption of a fixed spacetime background acting as a stage for the propagation of quantum information is at odds with the ``It from Qubit" paradigm, which seeks to reconstruct classical spacetime physics from the structure of quantum information. Moreover, the cross-disciplinary utility of quantum information theory has solidified the conceptual foundations of this paradigm in recent years, thus fueling a nascent approach to fundamental physics which views spacetime as emergent rather than fundamental~\cite{Bombelli_1987,Jacobson_1995,ambjorn_2004,Raamsdonk_2010,Swingle_2012,Arkani_2013,Gielen_2013,Cao_2016,FV25,Taka_25}.   

In this Letter, we take the view that if spacetime is indeed an emergent property of interacting qubits, then its symmetries should be encoded in the internal symmetries of qubits themselves, without any reference to external variables such as position and momentum. In accordance with such a `pre-spacetime' viewpoint, we then derive the action of the proper orthochronous Lorentz group $\text{SO}^+(1,3)$ on the internal degrees of freedom of a qubit from a simple, information-theoretic principle we refer to as \emph{preservation of linear entropy} for a single qubit. 

The linear entropy $S_L(\rho)$ of a density matrix $\rho$ is obtained by replacing $\ln(\rho)$ in the expression $-\Tr(\rho \ln(\rho))$ defining the von~Neumann entropy by its linear approximation $\rho-\mathds{1}$, thus resulting in the formula
\be \label{LINXRENT67}
S_L(\rho)=1-\Tr(\rho^2)\, .
\ee
Similar to von~Neumann entropy, the linear entropy may then be viewed as a measure of purity, thus the preservation of linear entropy principle we invoke in our derivation of the Lorentz transformations may also be thought of as a preservation of purity for a single qubit. 

If $\Lambda\in \text{SL}(2,\C)$ corresponds to a Lorentz boost under the spin homomorphism $\text{SL}(2,\C)\to \text{SO}^+(1,3)$, then the mapping $\rho\mapsto \Lambda \rho\Lambda^{\dag}$ is positive but not trace-preserving. As such, it is natural to work with un-normalized states in the context of relativistic quantum information. For $\rho$ a (normalized) density matrix representing the state of a single qubit, it turns out that 
\be \label{DXTX71}
S_L(\rho)=\Tr(\rho\rho^{\star})\, ,
\ee
where $\rho^{\star}=Y\overline{\rho}Y$ is the \emph{spin-flip} of $\rho$ (here $Y$ denotes the Pauli-$Y$ matrix and $\overline{\rho}$ denotes the complex conjugate of $\rho$). Taking formula \eqref{DXTX71} as the definition of linear entropy for possibly un-normalized single-qubit states, we obtain an $\text{SL}(2,\C)$ invariant scalar that may be viewed as a measure of the distinguishability between a state $\rho$ and its spin-flip $\rho^{\star}$. 

The notion of spin-flip naturally extends to $n$-qubit states $\rho$, and for $n=2$ the associated $W$-matrix given by $W=\rho\rho^{\star}$ was utilized by Hill and Wootters to define a measure of 2-qubit entanglement referred to as \emph{concurrence}~\cite{Hill_1997}. The concurrence is defined in terms of the eigenvalues of $W$, and it was later shown to be an $\text{SL}(2,\C)^{\otimes 2}$ invariant scalar in Ref.~\cite{Verstraete_2003}. Here we show that the Lorentz invariance of the concurrence for 2-qubit states and the linear entropy for single-qubit states are both manifestations of a much more general phenomenon, namely, that any spectral invariant of the $n$-qubit $W$-matrix is an $\text{SL}(2,\C)^{\otimes n}$ invariant scalar for any $n$-qubit density matrix $\rho$.

 We then prove a remarkable fact: If $W=\rho\rho^{\star}$ is the $W$-matrix associated with an $n$-qubit state $\rho$, then
 \be \label{RMARX87}
 \Tr(W)=I_L(\rho)\, ,
 \ee
where $I_L(\rho)$ is the \emph{linear $n$-partite quantum mutual information of $\rho$}, which is obtained from the $n$-partite quantum mutual information by replacing von~Neumann entropy with the linear entropy. Moreover, it follows from the $\text{SL}(2,\C)^{\otimes n}$ invariance of the spectral invariants of $W$ that the linear $n$-partite quantum mutual information is $\text{SL}(2,\C)^{\otimes n}$ invariant as well.

For the case $n=2$ it is known that $\Tr(W)=C^2$, where $C$ is the concurrence associated with a pair of qubits in a pure state $\rho$. It then follows from \eqref{RMARX87} that in such a case we have $C^2=I_L(\rho)$, identifying the concurrence of a pure state of 2-qubits as the square root of the linear bi-partite quantum mutual information. Therefore, the linear $n$-partite quantum mutual information is a natural, Lorentz invariant generalization of concurrence for all $n>2$.

Finally, in the Heisenberg picture where it is the observables which transform rather than the states, we show that the correlation function associated with a pair of qubits in the singlet state $\ket{\Psi^-}=(\ket{01}-\ket{10})/\sqrt{2}$ yields the Minkowski metric on the space of qubit observables, whose symmetry group is the full Lorentz group $\text{SO}(1,3)$. As such, the results in this Letter reveal the ubiquity of Lorentzian symmetries of quantum information in a pre-spacetime setting, in both the Schr\"{o}dinger and Heisenberg pictures of quantum theory.

\emph{Entropic derivation of the Lorentz transformations}. Throughout this Letter we let $\text{Herm}_2$ denote the real vector space of $2\times 2$ Hermitian matrices, and we let $\{\mathds{1},X,Y,Z\}$ denote the associated Pauli basis. In the spirit of Special Relativity, we view linear isomorphisms $T:\text{Herm}_2\to \text{Herm}_2$ as transition functions between equivalent descriptions of events in a pre-spacetime quantum substrate. Positive transition functions $T:\text{Herm}_2\to \text{Herm}_2$ may then be viewed as transformations of (possibly un-normalized) single-qubit states, which are positive elements $\rho\in \text{Herm}_2$ with $\Tr(\rho)>0$. A state $\rho$ of rank-1 will be referred to as \emph{pure}.

We now show that the group of completely positive transition functions $T:\text{Herm}_2\to \text{Herm}_2$ which preserve the linear entropy of states may be naturally identified with the restricted Lorentz group $\text{SO}^+(1,3)$. We recall that the linear entropy of a qubit in a (possibly un-normalized) state $\rho$ is the non-negative real number $S_L(\rho)$ given by
\[
S_L(\rho)=\Tr(\rho)^2-\Tr(\rho^2)\, .
\]

So now let $T:\text{Herm}_2\to \text{Herm}_2$ be a completely positive transition function which preserves linear entropy of states. As it is straightforward to show that $\rho$ is pure if and only if $S_L(\rho)=0$, the assumption of preservation of linear entropy implies that $T$ takes pure states to pure states. Moreover, as a state $\rho=t\mathds{1}+xX+yY+zZ$ is pure if and only if $\det(\rho)=t^2-x^2-y^2-z^2=0$, the fact that $T$ takes pure states to pure states together with the fact that $T$ is linear implies that the quadratic form 
\[
q(t,x,y,z)=\det\big(T(t\mathds{1}+xX+yY+zZ)\big)
\]
vanishes on the zero-locus of the quadratic form
\[
p(t,x,y,z)=t^2-x^2-y^2-z^2\, .
\]
Now since $p(t,x,y,z)$ is an irreducible polynomial over $\R$ which is indefinite (meaning it takes on both positive and negative values), it follows from the real Nullstellensatz that there exists a real number $\lambda\neq 0$ such that
\[
q(t,x,y,z)=\lambda p(t,x,y,z) \quad \forall (t,x,y,z)\in \R^{1,3}\, . 
\]  
We then conclude that there exists $\lambda\neq 0$ such that for all $\O\in \text{Herm}_2$, $\det(T(\O))=\lambda \det(\O)$. Moreover, since for all states $\rho\in \text{Herm}_2$ we have
\be \label{DETXF67}
S_L(\rho)=\Tr(\rho)^2-\Tr(\rho^2)=2\det(\rho)\, ,
\ee
our assumption that $T$ preserves the linear entropy of states $S_L(\rho)=S_L(T(\rho))$ together with the fact that  $\det(T(\O))=\lambda \det(\O)$ yields
\begin{align*}
2\det(\rho)&=S_L(\rho)=S_L(T(\rho)) \\
&=2\det(T(\rho))=2\lambda\det(\rho) \\
&\implies \lambda=1\, ,
\end{align*}
thus $\det(T(\O))=\det(\O)$ for all $\O\in \text{Herm}_2$.

Finally, since $T:\text{Herm}_2\to \text{Herm}_2$ is a completely positive linear map which preserves the determinant, it necessarily follows that there exists an element $\Lambda\in \text{SL}(2,\C)$ such that $T(\O)=\Lambda \O \Lambda^{\dag}$ for all $\O\in \text{Herm}_2$. Therefore, the preservation of linear entropy assumption implies that $T$ acts on $\text{Herm}_2$ via the spin homomorphism $\text{SL}(2,\C)\to \text{SO}^+(1,3)$, and thus may identified with a unique element of $\text{SO}^+(1,3)$. Furthermore, since it follows from Eqs.~\eqref{DETXF67} that any mapping of the form $\O\mapsto \Lambda \O \Lambda^{\dag}$ with $\Lambda\in \text{SL}(2,\C)$ necessarily preserves linear entropy, the group of completely positive transition functions which preserve linear entropy may be naturally identified with $\text{SO}^+(1,3)$.

\emph{Spectral invariants of the $W$-matrix}. The $W$-matrix of an $n$-qubit state $\rho\in \text{Herm}_2^{\otimes n}$ is the matrix $W=\rho\rho^{\star}$, where $\rho^{\star}=Y^{\otimes n}\overline{\rho}\,Y^{\otimes n}$ is the \emph{spin-flip} of $\rho$ (so that $\overline{\rho}$ the complex conjugate of $\rho$). For $n=2$ the eigenvalues of the $W$-matrix were utilized by Hill and Wootters to define the \emph{concurrence} of a 2-qubit state $\rho$~\cite{Hill_1997}, which is a fundamental measure of entanglement.

We now show that spectral invariants of the $W$-matrix associated with an $n$-qubit state $\rho$ are $\text{SL}(2,\C)^{\otimes n}$ invariant scalars for all $n>0$. For this, let $\rho\in \text{Herm}_2^{\otimes n}$ be an $n$-qubit state, let $\bold{Y}=Y^{\otimes n}$, let $\Lambda_i\in \text{SL}(2,\C)$ for $i=1,\ldots,n$, let $M=\Lambda_1\otimes \cdots \otimes \Lambda_n$, and let $\rho'=M\rho M^{\dag}$, so that
\begin{align*}
(\rho')^{\star}&=\bold{Y}\overline{\rho'}\bold{Y}=\bold{Y}\overline{M\rho M^{\dag}}\bold{Y} \\
&=\bold{Y}\overline{M}\overline{\rho} \overline{M^{\dag}}\bold{Y}=\bold{Y}\overline{M}\overline{\rho} M^T\bold{Y}\, .
\end{align*}
Now since $Y\overline{\Lambda}=\Lambda^{\vee}Y$ and $\Lambda^TY=Y\Lambda^{-1}$ for all $\Lambda\in \text{SL}(2,\C)$, where $\Lambda^{\vee}=(\Lambda^{\dag})^{-1}$, it follows that $\bold{Y}\overline{M}=M^{\vee}\bold{Y}$ and $M^T\bold{Y}=\bold{Y}M^{-1}$, thus
\[
(\rho')^{\star}=M^{\vee}\bold{Y}\overline{\rho}\bold{Y}M^{-1}=M^{\vee}\rho^{\star}M^{-1}\, .
\]
We then have
\begin{align*}
W'&=\rho'(\rho')^{\star}=(M\rho M^{\dag})(M^{\vee}\rho^{\star}M^{-1}) \\
&=M\rho\rho^{\star}M^{-1}=MWM^{-1}\, ,
\end{align*}
thus $W$ and $W'$ have the same spectrum. It then follows that any spectral invariant of the $W$-matrix is an $\text{SL}(2,\C)^{\otimes n}$ invariant scalar. 

Given a single qubit state $\rho\in \text{Herm}_2$, the spin-flip of $\rho$ may be written as $\rho^{\star}=\Tr(\rho)\mathds{1}-\rho$. Therefore, we have
\begin{align*}
\Tr(W)&=\Tr(\rho(\Tr(\rho)\mathds{1}-\rho)) \\
&=\Tr(\rho)^2-\Tr(\rho^2)=S_L(\rho)\, ,
\end{align*}
thus the $\text{SL}(2,\C)$ invariance of the linear entropy is a manifestation of the fact that spectral invariants of the $W$-matrix for $n=1$ are $\text{SL}(2,\C)$ invariant. Similarly, for $n=2$ the concurrence of $\rho$ is defined to be the non-negative quantity $\text{max}\{0,\sqrt{\lambda_1}-\sqrt{\lambda_2}-\sqrt{\lambda_3}-\sqrt{\lambda_4}\}$, where $\lambda_1\geq \lambda_2\geq \lambda_3\geq \lambda_4$ are the eigenvalues of $W$. It then follows that the $\text{SL}(2,\C)^{\otimes 2}$ invariance of concurrence (as shown in Ref.~\cite{Verstraete_2003}) is also a manifestation of the fact that spectral invariants of the $W$-matrix are Lorentzian symmetries of quantum information.

\emph{Linear $n$-partite quantum mutual information}. For an $n$-qubit state $\rho$, the linear $n$-partite quantum mutual information $I_L(\rho)$ is given by
\begin{equation} \label{ILX87}
I_L(\rho) = \sum_{A \subset \{1,\dots,n\}} (-1)^{|A|+1}\, S_L(\rho_A)\, ,
\end{equation}
where $\rho_A$ denotes the reduced density matrix associated with subsystem $A$. While the linear $n$-partite quantum mutual information is a mysterious quantity that has not received much attention in the literature, in the End Matter we prove a remarkable formula: $I_L(\rho)=\Tr(W)$, where $W=\rho\rho^{\star}$ is the associated $W$-matrix. Such a formula not only bypasses the combinatorial complexity of the defining formula \eqref{ILX87}, but it also reveals $I_L(\rho)$ as a global measure of distinguishability between $\rho$ and its spin flip $\rho^{\star}$. Moreover, as we have already established that spectral invariants of the $W$-matrix are $\text{SL}(2,\C)^{\otimes n}$ invariant, it follows that $I_L(\rho)$ is $\text{SL}(2,\C)^{\otimes n}$ invariant for all $n$-qubit states $\rho$.

Other direct implications of the trace formula $I_L(\rho)=\Tr(W)$ include the following: First, since for qubits in a product state $\rho_1\otimes \rho_2 \otimes \dots \otimes \rho_k$ the $W$-matrix is given by $W=\rho_1\rho_1^\star\otimes \rho_2\rho_2^\star \otimes \dots \otimes \rho_k\rho_k^\star$, the trace formula $I_L(\rho)=\Tr(W)$ immediately yields the multiplicative property
\[
I_L(\rho_1\otimes \cdots \otimes \rho_k) = \prod_{i=1}^k I_L(\rho_i)\, .
\]
Second, since the $W$-matrix has the same eigenvalues as the positive matrix $\sqrt{\rho}\rho^{\star}\sqrt{\rho}$, the trace formula $I_L(\rho)=\Tr(W)$ implies that $I_L(\rho)\geq 0$ for all $n$-qubit states $\rho$. Third, since the concurrence $C$ associated with a $2$-qubit pure state $\rho$ coincides with $\sqrt{2S_L(\rho_1)}$ (where $\rho_1$ is a reduced density matrix of $\rho$), the vanishing of linear entropy for pure states together with the trace formula $I_L(\rho)=\Tr(W)$ implies $I_L(\rho)=C^2$. As such, the linear $n$-partite quantum mutual information may be viewed as a generalization of concurrence to $n$-qubit states for all $n>2$.

However, $I_L$ is not a faithful entanglement measure for odd $n$, since in such a case $I_L$
vanishes identically on all pure states. This is due to the fact that an odd number of qubits cannot be decomposed into a union of spin-flipped pairs, resulting in a pure state being distinguishable from its spin-flipped counterpart. This also explains why concurrence---when viewed as an entanglement measure as opposed to linear quantum mutual information---does not generalize to triples of qubits. Nevertheless, while the standard tripartite quantum mutual information can take negative values in the presence of synergistic correlations~\cite{Kitaev_2006,Hayden_2011,Sesh_2018}, its linear analog remains strictly non-negative. 

For even $n$, the linear $n$-partite quantum mutual information serves as a nontrivial monotone capable of detecting multipartite correlations. For $n=4$, $I_L$ vanishes on $\mathcal{W}$-type states while remaining non-vanishing for both the GHZ state and the tensor product of two singlets~\cite{Wong_2001}. This distinction highlights its sensitivity to correlation structures fundamentally different from those in the $\mathcal{W}$ class. While a precise physical interpretation of $I_L$ remains to be elucidated, our results suggest that it provides an information-theoretic primitive for investigating the entropic origins of relativistic symmetries.

\emph{The correlation function of the singlet state}. The singlet state $\ket{\Psi^-}=(\ket{01}-\ket{10})/\sqrt{2}$ is the most remarkable state of a pair of qubits. Not only is it maximally entangled, it is also maximally symmetric in the sense that $\Lambda\otimes \Lambda\ket{\Psi^-}=\det(\Lambda)\ket{\Psi^-}$ for all $\Lambda\in \text{Mat}(2,\C)$. This maximal symmetry implies that the correlation function of the singlet $\mathscr{C}(\O_1,\O_2)=\bra{\Psi^-}\O_1\otimes \O_2\ket{\Psi^-}$ is such that
\[
\mathscr{C}(\O_1,\O_2)=\mathscr{C}(\Lambda \O_1 \Lambda^{\dag},\Lambda \O_2 \Lambda^{\dag}) \quad \forall \Lambda, |\det(\Lambda)|=1\, .
\]
As every $U\in \text{U}(2)$ satisfies $|\det(U)|=1$, it follows that
\[
\mathscr{C}(\O_1,\O_2)=\int_{\mathrm{U}(2)} \mathscr{C}(\Ad_U(\O_1),\Ad_U(\O_2)) \, \mathrm{d}\mu\, ,
\]
where $\mu$ is the Haar measure on $\text{U}(2)$ and $\Ad_U(\O)=U\O U^{\dag}$ for all $\O\in \text{Herm}_2$. By results on integrals over unitary groups~\cite{zhang2014}, it follows that
\[
\int_{\mathrm{U}(2)} \text{Ad}_{U\otimes U}(\O_1\otimes \O_2) \, \mathrm{d}\mu=\chi(\O_1,\O_2)\mathds{1}- \zeta(\O_1,\O_2) F\, ,
\]
where $F$ is the swap operator,
\[
\chi(\O_1,\O_2)=\left( \frac{\Tr(\O_1) \Tr(\O_2)}{3} - \frac{\Tr(\O_1 \O_2)}{6} \right)\, ,
\]
and
\[
\zeta(\O_1,\O_2)=\left( \frac{\Tr(\O_1)\Tr(\O_2)}{6} - \frac{\Tr(\O_1 \O_2) }{3} \right)\, .
\]
We then have
\begin{align*}
\mathscr{C}(\O_1,\O_2)&=
\chi(\O_1,\O_2)\braket{\Psi^-}{\Psi^-}
- \zeta(\O_1,\O_2)\bra{\Psi^-}F\ket{\Psi^-}
\\
&=
\left( \frac{\Tr(\mathscr{O}_1) \Tr(\mathscr{O}_2)}{3} - \frac{\Tr(\mathscr{O}_1 \mathscr{O}_2)}{6} \right) \\
&\hspace{0.5cm}- \left( \frac{\Tr(\mathscr{O}_1)\Tr(\mathscr{O}_2)}{6} -
\frac{\Tr(\mathscr{O}_1 \mathscr{O}_2) }{3} \right)(-1) \\
&=\frac{1}{2}\big(\Tr(\mathscr{O}_1) \Tr(\mathscr{O}_2)-\Tr(\mathscr{O}_1 \mathscr{O}_2)\big) \\
&=\frac{1}{2}\big(\det(\O_1+\O_2)-\det(\O_1)-\det(\O_2)\big)\, ,
\end{align*}
identifying $\mathscr{C}$ as the symmetric bilinear form obtained by polarizing the determinant. 

It then follows that $\mathscr{C}(\O_1,\O_2)=\mathscr{C}(L(\O_1),L(\O_2))$ for any linear map $L:\text{Herm}_2\to \text{Herm}_2$ which preserves the determinant, establishing the full Lorentz group $\text{SO}(1,3)$ as the group of symmetries of the correlation function of the singlet state.  Moreover, as expanding an observable $\O=t\mathds{1}+xX+yY+zZ$ with respect to the Pauli basis yields
\[
\mathscr{C}(\O,\O)=\det(\O)=t^2-x^2-y^2-z^2\, ,
\]
one finds that the correlation function of the singlet state is precisely the Minkowski metric on the space of qubit observables $\text{Herm}_2$. 

The fact that the structure of correlations of EPR pairs in a singlet state endows the space of qubit observables with the Minkowski metric provides compelling evidence in favor of the ``It from Qubit" paradigm. As such, it is tempting to speculate that the mapping
\[
\left(
\begin{array}{cc}
t+z&x-iy\\
x+iy&t-z
\end{array}
\right)\longmapsto (t,x,y,z)
\]
taking a $2\times 2$ Hermitian matrix to its associated Minkowski 4-vector is not merely a mathematical correspondence~\cite{Arrighi_2003}, but a mathematical \emph{description} of how spacetime emerges from quantum information. However, the formulation of an explicit \emph{mechanism} for such an emergence of spacetime from quantum information remains an open and profound challenge.

\emph{Concluding remarks}. While it was shown by Peres, Scudo, and Terno that von Neumann entropy has no invariant meaning when the Lorentz group acts on the momentum degrees of freedom of a system of qubits embedded in a fixed spacetime background~\cite{Peres_2002}, our results establish the ubiquity of Lorentzian symmetries of quantum information. Specifically, we have shown that spectral invariants of the $W$-matrix serve as Lorentzian invariants of quantum information in a pre-spacetime setting where only the internal degrees of freedom of a qubit are considered. Special cases of this result include the $\text{SL}(2,\C)$ invariance of linear entropy for qubit states and the $\text{SL}(2,\C)^{\otimes 2}$ invariance of concurrence for pairs of qubits. We then showed for $n$-qubit states $\rho$ that the trace of the associated $W$-matrix coincides with the linear $n$-partite quantum mutual information of $\rho$, establishing the linear $n$-partite mutual information as an $\text{SL}(2,\C)^{\otimes n}$ invariant for all $n>0$. 

Furthermore, in the spirit of the ``It from Qubit" paradigm, we \emph{derived} the action of the restricted Lorentz group $\text{SO}^+(1,3)$ on the internal degrees of freedom of a qubit from the simple, information-theoretic assumption of preservation of linear entropy, thus providing a information-theoretic derivation of the Lorentz transformations from a purely quantum foundation. Such results suggest that the apparent conflict between relativity and quantum information is not an intrinsic feature of quantum information, but rather a limitation of the background-dependent framework in which it is traditionally described.

Of course the question remains as to how gravity should be incorporated into the pre-spacetime framework developed in this Letter. As the restricted Lorentz group $\text{SO}^+(1,3)$ acting on single qubit states preserves the linear entropy---and hence purity---it is natural to surmise that gravitational effects are fundamentally linked with the onset of decoherence. Such a perspective aligns with theories of ``gravitationally induced decoherence'' such as the Penrose-Diósi proposal~\cite{Diosi_1989, Penrose_1996}, wherein gravitational effects mediate the quantum-to-classical transition. Reconciling our pre-spacetime framework with the notion of gravitational decoherence may then offer a formal path toward understanding how the geometry of curved spacetime may emerge from the structure of quantum information.

\emph{Acknowledgments.}  JF is supported by the Key Development Project of Hainan Province for the project ``Spacetime from Quantum Information", grant no. 126MS0010. VV acknowledges support from the Templeton and the Gordon and Betty Moore Foundations.
EG-G is supported by Hainan Provincial Natural Science Foundation of China, grant no. 126MS0008.

\bibliography{references}

\appendix
\onecolumngrid
\section*{End Matter}
\twocolumngrid
\section{The trace formula for linear $n$-partite quantum mutual information} \label{app:B}

In this Appendix we prove the trace formula $I_L(\rho)=\Tr(W)$, where $I_L$ is the linear $n$-partite quantum mutual information and $W=\rho\rho^{\star}$ is the $W$-matrix associated with an $n$-qubit state $\rho$. For this, consider an arbitrary, possibly un-normalized state $\rho$ of a system of $n$ qubits. When expressed in a
product basis of Hermitian operators, it can be written as
\[
\rho = A_i^1 \otimes A_i^2 \otimes \cdots \otimes A_i^n\, ,
\]
where $A_i^m \in \mathrm{Herm}_2$ acts on the $m$-th qubit, and a sum over repeated indices is implicit.

After tracing out subsystems $\alpha_1 < \dots < \alpha_q$ while retaining $\beta_1 < \dots < \beta_p$, we obtain
\begin{equation}
\rho_{\beta_1 \dots \beta_p}
= \Tr(A_i^{\alpha_1}) \cdots \Tr(A_i^{\alpha_q})\,
A_i^{\beta_1} \otimes \cdots \otimes A_i^{\beta_p}\, .
\label{eq:rhobeta}
\end{equation}
Since the spin-flip operator acts locally, we have
$\rho^\star = A_j^1{}^\star \otimes A_j^2{}^\star \otimes \cdots \otimes A_j^n{}^\star$,
thus
\begin{equation} \label{AXS27}
\begin{split}
\Tr(\rho \rho^\star)
&= \Tr\big(
 A_i^1A_j^1{}^\star \otimes A_i^2A_j^2{}^\star \otimes \cdots \otimes A_i^nA_j^n{}^\star
 \big)
\\
&= \Tr(A_i^1A_j^1{}^\star)\,\Tr(A_i^2A_j^2{}^\star)\cdots \Tr(A_i^nA_j^n{}^\star)\, .
\end{split}
\end{equation}
Since the spin-flip satisfies $\rho^\star = \Tr(\rho)\,\mathds{1} - \rho$ for single-qubit operators, it follows that
$
\Tr(\rho_1 \rho_2^\star)
= \Tr(\rho_1)\Tr(\rho_2) - \Tr(\rho_1 \rho_2)
$.
Applying this result to each factor in \eqref{AXS27} then yields
\[
\Tr(\rho \rho^\star)
=\prod_{i=1}^{n} \big(\Tr( A_i^1 ) \Tr(A_j^1)- \Tr( A_i^1 A_j^1)\big) \, .
\]

When expanding the previous product, each factor contributes either a term of the
form $\Tr(A_i^m)\Tr(A_j^m)$ or $-\Tr(A_i^m A_j^m)$. Consider a choice of indices
$\alpha_1 < \dots < \alpha_q$ for which the first type is selected, and $\beta_1 < \dots < \beta_p$ for
which the second type is selected. Up to an overall factor $(-1)^p$, the
corresponding contribution is
\begin{equation}
\begin{split}
&\Tr( A_i^{\alpha_1})\Tr( A_j^{\alpha_1})\cdots
 \Tr( A_i^{\alpha_q})\Tr( A_j^{\alpha_q}) \\
&\qquad\times\,
 \Tr( A_i^{\beta_1} A_j^{\beta_1})\cdots
 \Tr( A_i^{\beta_p} A_j^{\beta_p})
\\
&=\Tr( A_i^{\alpha_1})\Tr( A_j^{\alpha_1})\cdots
  \Tr( A_i^{\alpha_q})\Tr( A_j^{\alpha_q}) \\
&\qquad\times\,
 \Tr\bigl( A_i^{\beta_1} A_j^{\beta_1}\otimes\cdots\otimes
            A_i^{\beta_p} A_j^{\beta_p}\bigr)
\\
&=\Tr( A_i^{\alpha_1})\Tr( A_j^{\alpha_1})\cdots
  \Tr( A_i^{\alpha_q})\Tr( A_j^{\alpha_q}) \\
&\qquad\times\,
 \Tr\bigl[(A_i^{\beta_1}\otimes\cdots\otimes A_i^{\beta_p})
          (A_j^{\beta_1}\otimes\cdots\otimes A_j^{\beta_p})\bigr]
\\
&=\Tr\Bigl(
\Tr( A_i^{\alpha_1})\cdots\Tr( A_i^{\alpha_q})\,
 A_i^{\beta_1}\otimes\cdots\otimes A_i^{\beta_p} \\
&\qquad\qquad\;
\Tr( A_j^{\alpha_1})\cdots\Tr( A_j^{\alpha_q})\,
 A_j^{\beta_1}\otimes\cdots\otimes A_j^{\beta_p}
\Bigr)
\\
&=\Tr(\rho_{\beta_1\cdots\beta_p}^2).
\end{split}
\label{eq:expansionSubsets}
\end{equation}
where we used (\ref{eq:rhobeta}) for the last equality.

By summing over all choices of $\beta_1 < \dots < \beta_p$, i.e., over all subsets
$\bm{\beta} \subset \{1,\dots,n\}$ (with $\{\alpha_1,\dots,\alpha_q\}$ given by the complement of $\bm{\beta}$), we
obtain
\begin{equation}
\Tr(\rho \rho^\star)= \sum_{\bm{\beta}} (-1)^{|\bm{\beta}|} \Tr(\rho_{\bm{\beta}}^2)\, ,
\label{eq:rhorhostar0}
\end{equation}
where $|\bm \beta|$ denotes the cardinality of $\bm{\beta}$.

Finally, note that $\Tr(\rho_{\bm{\beta}})=\Tr(\rho)$ for all subsets $\bm{\beta}$, and
$\sum_{\bm{\beta}} (-1)^{|\bm{\beta}|}=0$, which follows by an argument analogous to the
expansion leading to Eq.~(\ref{eq:expansionSubsets}). Therefore,
\[
\sum_{\bm{\beta}} (-1)^{|\bm{\beta}|} \Tr(\rho_{\bm{\beta}})^2
= \Tr(\rho)^2 \sum_{\bm{\beta}} (-1)^{|\bm{\beta}|}
= 0\, .
\]
Subtracting $0=\Tr(\rho)^2 \sum_{\bm{\beta}} (-1)^{|\bm{\beta}|}$ from (\ref{eq:rhorhostar0}), we obtain
\begin{align*}
\Tr(\rho \rho^\star)
&= \sum_{\bm{\beta}} (-1)^{|\bm{\beta}|} \Bigl(\Tr(\rho_{\bm{\beta}}^2) - \Tr(\rho_{\bm{\beta}})^2\Bigr) \\
&= \sum_{\bm{\beta}} (-1)^{|\bm{\beta}|+1} S_L(\rho_{\bm{\beta}}) \\
&=I_L(\rho)\, ,
\end{align*}
as desired.

\clearpage

\end{document}